\input epsf
\input cp-aa
\def\etal{et al.} 
\def\PA{{PA}} 
\def\PAm{{\hbox{PA}}} 
\newcount\fignumber\fignumber=1 
\def\nfig{\global\advance\fignumber by 1} 
\def\fignam#1{\xdef#1{\the\fignumber}} 
\def\nomps#1{\vbox{\epsfbox{#1}\vfil}} 
 
\fignam{\FPROF}   \nfig 
\fignam{\FBETH}   \nfig 
\fignam{\FBEGA}   \nfig 
\fignam{\FEPSI}   \nfig 
\fignam{\FTYPE}   \nfig 
\fignam{\FDEPRO}  \nfig 
\fignam{\FNSTIM}  \nfig 
\fignam{\FNSCOM}  \nfig 
\fignam{\FNSORI}  \nfig 
\def\tabsample{1} 
\def\tabresult{2} 
\MAINTITLE={Disc galaxies with multiple triaxial structures 
\FOOTNOTE{Based on observations collected at European Southern 
Observatory and Steward Observatory}} 
\SUBTITLE={II. JHK surface photometry and numerical simulations} 
\AUTHOR={D. Friedli@{1}, H. Wozniak@{2}, M. Rieke@{3}, 
         L. Martinet@{1} and P. Bratschi@{1}}                                
\INSTITUTE={ 
@1 Observatoire de Gen\`eve, CH-1290 Sauverny, Switzerland 
@2 Observatoire de Marseille, F-13248 Marseille cedex 4, France  
@3 Steward Observatory, University of Arizona, Tucson AZ 85721, USA} 
\OFFPRINTS={D. Friedli, fri\at scsun.unige.ch} 
\DATE={Received 16 October 1995, accepted 2 February 1996} 
 
\ABSTRACT={We present detailed JHK surface photometry with ellipse  
fits of 13 galaxies selected from previous optical observations as
likely candidates for having a secondary bar or a triaxial bulge
within the primary bar.  We have found 7 double-barred galaxies, 3
double-barred galaxies with an additional intermediate structure with
twisted isophotes, and 3 galaxies with a bar and central twisted
isophotes.  A global analysis of the structural parameter
characteristics in the I- and K-bands is presented.  Various numerical
models of galaxies with bars within bars are also analysed using the
ellipse fitting technique and compared to the observations. A thorough
review of the possible hypotheses able to explain this phenomenon is
given with emphasis on the most likely ones.}
 
\KEYWORDS={Galaxies: photometry -- Galaxies: structure -- Galaxies:  
fundamental parameters -- Galaxies: nuclei -- Galaxies: evolution -- 
Galaxies: barred} 
\THESAURUS={11(11.16.1; 11.19.6; 11.06.2; 11.14.1; 11.05.2)} 
\maketitle 
 
\AUTHORRUNNINGHEAD{Friedli \etal} 
\MAINTITLERUNNINGHEAD{Disc galaxies with multiple triaxial structures. II} 
 
\titlea{Introduction} 
In the first paper of this series (Wozniak \etal\ 1995, hereafter
Paper I), we have presented a BVRI and H$\alpha$ survey of disc
galaxies having multiple triaxial structures.  More than one triaxial
structure has been found in 22 galaxies, i.e. these galaxies show,
inside the large-scale primary bar, at least one misaligned secondary
bar or one structure with twisted isophotes.
 
Here, we present the detailed analysis of JHK surface photometry for a
subsample of this survey.  An extended and accurate knowledge of the
structural parameters of observed single or multiple bars is a
prerequisite to thorough comparisons with numerical simulations, and
should give constraints and clues on the various processes of galactic
secular evolution.  Different kinds of information can be inferred
from various bands, in particular the K-band traces fairly well the
luminosity of the old stellar population (e.g. Rix \& Rieke 1993; but
see also Rhoads 1995) which thus opens the door to a better knowledge
of the dynamical mass distribution (except dark matter).  So far,
published near-infrared surface photometry observations of large
samples of galaxies have unfortunately not been very numerous
(Baumgart \& Peterson 1986; Terndrup \etal\ 1994; de Jong \& van der
Kruit 1994).  Independently of the central problem of this paper, a
significant increase of this type of data is certainly most desirable
and necessary.
 
Since the paper of de Vaucouleurs (1974) more than twenty years ago,
more and more compelling evidence has been accumulated that not only
the majority of galaxies are barred (e.g. Sellwood \& Wilkinson 1993),
but also that some of them host at least two misaligned nested bars or
one bar plus one triaxial bulge (e.g. Kormendy 1979, 1982ab; Jarvis
\etal\ 1988; Friedli \& Martinet 1993; Buta \& Crocker 1993; Shaw 
\etal\ 1993, 1995; Paper I; Friedli 1996).  These central isophotal 
deformations have thus been known for a long time and appear to be a 
widespread feature of disc galaxies although no reliable percentage 
can yet be given. The nuclear isophotes are subject to various 
asymmetries and twists resulting in a very complex morphology which 
indicates an intricate dynamics as well.  However, the presence of 
dust confuses considerably the picture in optical bands.  Since in the 
near-IR, especially in the K-band, the problems of dust absorption are 
greatly minimized (only about one-tenth of that at visible 
wavelengths), the existence and the shape of the various triaxial 
structures can be determined much more firmly. 
 
The organization of the paper is the following: Sect.~2 is devoted to
the description of our sample and reduction processes.  In Sect.~3, we
present an individual description of galaxies, the global properties
of galaxies with multiple triaxial structures, and a comparison with
what was found with BVRI (Paper I).  An analysis of various 3D
numerical simulations with stars, gas and star formation is given in
Sect.~4.  A discussion is presented in Sect.~5 and our conclusions are
summarized in Sect.~6.
 
\titlea{The observations} 
Our sample is presented in Table~{\tabsample}. These galaxies
represent a subsample of the ones analysed in Paper I.  Observations
were taken in both hemispheres and they are presented below in turn.
 
\begtabfull 
\tabcap{\tabsample}{ 
Sample of observed galaxies} 
\rm{
\vbox{\tabskip=0pt
\halign to \hsize{
#
&\hskip2truemm #\hfill &\hskip2truemm \hfill# 
&\hskip2truemm \hfill# &\hskip2truemm \hfill#\hfill
&\hskip2truemm \hfill#\hfill 
& # \tabskip=0pt \cr
\noalign{\hrule\smallskip}
& \hfill Names \hfill & \hfill PA$_{\rm disc}$ \hfill 
& \hfill $i$ \hfill & \hfill References \hfill
& \hfill Observations \hfill &\cr
\noalign{\smallskip}
& \hfill (1) \hfill & \hfill (2) \hfill 
& \hfill (3) \hfill & \hfill (4) \hfill & \hfill (5) \hfill\cr
\noalign{\smallskip\hrule\smallskip}
& NGC~470    &  149$\pm$3 &  51$\pm$4 & a & E2.2 &\cr
& NGC~1097   &  134       &  46$\pm$5 & b & E2.2 &\cr
& NGC~2681   &   50       &  32       & f & S1.5 &\cr
& NGC~2950   &  110       &  50       & f & S1.5 &\cr	
& NGC~3081   &  123       &  33       & f & S1.5 &\cr
& NGC~4314   &  121$\pm$10 &  30$\pm$5 & c & S1.5 &\cr
& NGC~4340   &   85       &  47       & f & S1.5 &\cr
& NGC~5850   &  157       &  37       & f & S1.5 &\cr
& NGC~5905   &   45       &  40       & d & S1.5 &\cr
& NGC~6782   &   40       &  26       & f & E2.2 &\cr
& NGC~6951   & 135$\pm$10 &  28       & e & S2.3 &\cr
& NGC~7098   &   50       &  66       & f & E2.2 &\cr
& NGC~7479   &   37$\pm$2 &  44$\pm$2 & a & S2.3 &\cr
 \noalign{\smallskip\hrule\smallskip}
}}
\parindent=5truemm
{\advance\leftskip by \parindent 
\parindent=-\leftskip

Columns (2) and (3): Published Position-Angle and inclination of the
disc in degrees

Column (4): References for Columns (2) and (3):
a = Garcia G\`omez \& Athanassoula (1991) / 
b = Ondrechen \etal\ (1989) /
c = Wakamatsu \& Nishida (1980) / 
d = van Moorsel (1982) /
e = Boer \& Schulz (1993) /
f = our measurements

Column (5): E2.2 = observed at ESO 2.2m at La Silla / S1.5 = observed
at Steward Observatory 1.5m at Mt.~Bigelow / S2.3 = observed at
Steward Observatory 2.3m at Kitt Peak

}}
 
\endtab 
 
\titleb{ESO sample} 
The observations were carried out during a run of two nights at ESO
with the 2.2m telescope at La Silla. Photometric conditions were only
realized during the first part of the first night and the second
night.  The photometry for NGC~1097 and NGC~6782 (especially the
K-band) are thus not reliable. Over both nights and after complete
reduction (co-added exposures), the FWHM seeing was between
1.1\arcsec\ and 1.5\arcsec. On single exposure, it was between
0.9\arcsec\ and 1.2\arcsec. The detector (IRAC2 camera) was a
256$\times$256 NICMOS3 array.  The pixel size is 0.49\arcsec; the
field of view is thus about 2.1\arcmin\ large. Three galaxies
(NGC~1097, NGC~6782, NGC~7098) extend beyond this field of view. The
set of filters was chosen to be the standard JHK Johnson; we did not
use the K' filter.
 
For each object, we took 2 series of 4 exposures on the object, galaxy
or calibration star (science-frames), interspersed with sky exposures
(sky-frames). The integration time was 150~seconds giving a total of
20~minutes per filter and per object. Each frame was bias subtracted
and cleaned from cosmic rays as well as from cold and hot pixels.
Flat-field exposures were obtained from exposures on a uniform
illuminated blank screen. Flat-fields taken on the sky have been
constructed for each filter by normalizing all sky-frames and
averaging them using the median value of each coinciding pixels. These
sky-flats were compared with the dome-flats; they show significant
differences (cf. Moorwood
\etal\ 1992). We achieve a better flat background with dome-flats than
sky-flats.  Then, each science-frame and sky-frame were divided by the
flat-field. The sky background was removed using the mean of the two
sky-frames taken just before and just after the science-frame, except
for the first of the series for which there is only a sky exposure
taken after it.
 
As the telescope is moved between the science and sky images, the
exact position of the galaxy generally differs from one frame to
another and some regions are not 8 times exposed. However, no
correction was applied to these regions so that the effective field of
view is less than 2.1\arcmin\ large. A scaling of the flux in such
under-exposed region does not permit to achieve the same signal to
noise ratio as the noise is also amplified. Thus, we prefer to
restrict ourselves to the regions of good exposure.
 
Photometric calibrations were achieved by observing several
AASO/ESO/ISO standard stars 3 times during the night. Our precision on
the zero magnitude constant is 0.1 mag.(\arcsec)$^{-2}$ except for
NGC~1097 and NGC~6782.
 
\titleb{Steward Observatory sample} 
The observations were carried out during two runs of one and four
nights with the Steward Observatory 2.3m telescope on Kitt Peak and
the 1.5m telescope on Mt.~Bigelow.  Photometric conditions were
generally realized with a FWHM seeing of about 1.5\arcsec\ at Kitt
Peak and between 1.5\arcsec\ and 1.7\arcsec\ at Mt.~Bigelow.  Clouds
were however present during the observations of the H-image of
NGC~3081.  The detector used was a 256$\times$256 NICMOS3 array.  At
Kitt Peak, the pixel size was 0.6\arcsec\ corresponding to a field of
view of about 2.5\arcmin\ large, while at Mt.~Bigelow the pixel size
was 0.9\arcsec\ and the field of view about 3.8\arcmin.  Hereafter we
will refer to the K-band although the filter used in both runs was in
fact the K$_s$ filter with a center wavelength of
2.16$\mu$m. Individual exposures were typically of 30 or 60~seconds
with each galaxy being observed for a total of 40~minutes per filter
at the 1.5m and 20~minutes per filter at the 2.3m.
 
The data were reduced by first subtracting dark-frames from the
sky-frames. The skies were then median combined to create a
sky-flat. The average of the surrounding sky-images were subtracted
from each galaxy-frame which was then divided by the sky-flat. Images
were registered and shifted to a common center before the final median
combine.  We encountered saturation problems with the H-band of
NGC~2950 and K-band of NGC~4340. Indeed, the seeing was so good that
more flux hit on the central pixels than usual. The array response is
thus non-linear. Moreover, few of these pixels have a low sensitivity
(bad pixels). This give a saturated-like shape at the surface
brightness profiles.
 
Photometric calibrations were based on observations of Elias \etal\
(1982) standards. The photometric calibration is accurate to about 4\%
in flux based on the scatter of the data for individual standards.
For the two galaxies observed under non-photometric conditions, our
fluxes should not be trusted to any better than 30\%.
 
\titleb{Surface photometry analysis} 
We refer the reader to Paper I concerning the details of the technique
of ellipse fitting, and the terminology we are using. The various
effects which may affect the interpretation of the ellipticity $e$ and
the position-angle \PA\ have also extensively been discussed in Paper
I.

The following procedures have been performed to obtain the various
colour maps.  We have first aligned one image with respect to the
other. The transformation was obtained by measuring positions of
coinciding stars in the field of view. During the alignment, a scaling
factor was introduced in order to get the same pixel scale in both
bands.  After the rebinning at the same scale, the differences of
effective seeing between the images in different bands range from
0.05\arcsec\ to 0.1\arcsec. This is similar to the centering accuracy
when single exposure frames are co-added. We have however checked that
this small difference does not introduce any artifacts. To do this, we
have smoothed both images with a Gaussian filter to degrade the seeing
to the same value. The resulting colour maps share the same features
as the non-smoothed ones. Only the colour gradients are very slightly
changed whereas the morphological features for bulges and bars are
robust.  As our signal-to-noise ratio is too low in any event in the
discs to discuss their colours, we have chosen to display in
Figs.~{\FPROF} the non-smoothed colour maps.

Let us also briefly recall the main definitions.  The {\it primary
bar} is the bar with the larger spatial extent $l_p$ whereas the {\it
secondary bar} is the bar with the smaller spatial extent $l_s$.  The
angle between the two bars is $\theta$.  Positive values are for
leading secondary bars whereas negative values are for trailing
secondary bars (with regard to the primary bar rotation).  The length
ratio of the primary to the secondary bar is $\beta \equiv l_p/l_s$.
The integrated luminosity ratio between the two bars is $\gamma \equiv
L_p/L_s$ where $L_p$ and $L_s$ are the luminosity of the primary and
secondary bar.  Note that $L_p$ is defined as the luminosity inside
the isophote of semi-major axis length $l_p$ so that it includes the
luminosity of the secondary bar, the bulge and the nucleus. The same
definition holds for $L_s$.  The maximum ellipticities of the
secondary and primary bar are $e_s^{\rm max}$ and $e_p^{\rm max}$.  In
cases where more than two triaxial structures are present, they are
simply numbered from 1 to $n$ starting from the largest to the
shortest component. So, the semi-major axis length ratio of the
component $i$ to the component $j$ is $\beta_{ij}$ whereas the
integrated luminosity ratio is $\gamma_{ij}$.
 
\titlea{Results} 
\titleb{Individual description of galaxies} 
We give below a description of the main features observed for each
galaxy.  In general, the various parameters mentioned are always
measured on the K-image because this band is the one least affected by
dust extinction amongst JHK, and it best traces the old population and
hence mass.  At the end of the paper, Figs.~{\FPROF} show for each
galaxy in our sample the surface brightness $\mu$, ellipticity and
\PA\ profiles for JHK-filters, the greyscale and contour maps in J-
and K-bands as well as the greyscale map of the J--K colour. For a few
galaxies other bands or J--H or H--K colour maps are displayed.
 
\titled{NGC~470.}  
In Paper I we have detected two triaxial structures in this galaxy
usually classified SA(rs)b.  However, this was one of the most
difficult case to analyse since this galaxy contains a lot of
dust. Although we confirm here the presence of one primary bar, the
secondary component could be a bulge with the same orientation angle
as the disc. The deprojection is now much more difficult as the disc
appears warped (NGC~470 is in weak interaction with NGC~474). It is
thus questionable to assume that the outermost isophotes of the disc
are circular. The most sensible interpretation is that of a triaxial
bulge as the \PA\ profiles do not show any plateau but are twisted in
the innermost regions.  This central structure is clearly visible in
the J--K colour map as a very red component ($1.3 \la \rm J-K \la
1.6$).
 
\titled{NGC~1097.}
Our BVRI profiles obtained in Paper I were similar to those obtained
in the K-band by Shaw \etal\ (1993) and clearly indicated the presence
of two bars separated by a ring. However, the present data in
JHK-bands give a completely new view of the nuclear region. Indeed,
the circumnuclear ring is not closed.  Like Quillen \etal\ (1995), we
have found a small secondary bar ended by a trailing spiral-like
structure which seems to be an extension of the dust lanes.  The
isophote twists just outside the end of the secondary bar are due to
this spiral-like structure.
The resolution of Shaw \etal\ (1993) (1.24\arcsec/pixel) did not allow
them to clearly detect this peculiar structure although the isophote
shape of their Fig.~1a suggests some spiral structure at the secondary
bar end.  Knapen \etal\ (1995) have found a similar feature in
NGC~4321 but in that case additional leading spiral arms are also
present.  Dust prevents the detection of the secondary bar on HST
V-images (Barth \etal\ 1995).

The nature of the spiral-like structure is yet unclear. Is it a ring
crossed by two dust lanes in such a way that it appears not closed or
does the star formation occur on an actual spiral density wave (giving
rise to a ``hotspot'' look to the ring in the K-band)?  The western
dust lane, interpreted by Quillen \etal\ (1995) to be on the near side
of the galaxy, appears on the colour map but is more difficult to see
on the K-image.  Moreover, the spiral-like pattern has also been
observed in mid-infrared (Telesco \etal\ 1993) and radio observations
(Hummel \etal\ 1987).  However, as the dust is also visible on the
K-band in the region of the primary bar, we cannot exclude that, even
at 2.2$\mu$m, isophotes could be distorted by the darkening.  Note
that due to poor weather conditions during the observations, we are
not able to display any colour maps in Fig.~\FPROF.
 
\titled{NGC~2681.} 
It is difficult to identify unambiguously a secondary bar in this
galaxy, and the abundant dust seen in our BVRI-images could be
responsible for the apparent nuclear structure. Indeed, in the K-band,
the innermost ellipticity maximum is strongly decreased although still
present.  The primary bar shows a significant isophote twisting
(15\degr -- 20\degr\ depending on the band).  The J--K colour map
shows a blue Seyfert nucleus (as in Paper I) whose J--K is roughly
0.5~mag. bluer than the surrounding region. As the size of this nucleus
is closed to the seeing value, we have displayed the colour map
obtained from convolved images with approximately 3\arcsec.
 
\begtabfullwid 
\tabcap{\tabresult}{ 
Projected structural parameters of the various components for the 
K-band} 
\font\smallrm=cmr8

\font\smallit=cmti8
\smallrm{
\vbox{
\tabskip=0pt
\halign to \hsize{
#								
& #\hfill							
& \tabskip=.9pt plus.5pt minus.5pt \hskip1truemm #\hfil 	
& \hfill#\hfill							
& \tabskip=.9pt plus.5pt minus.5pt \hskip1truemm #\hfil 	
& \hfill# & \hfill# & \hfill#					
& \tabskip=.9pt plus.5pt minus.5pt \hskip1truemm #\hfil		
& \hfill# & \hfill# & \hfill#					
& \tabskip=.9pt plus.5pt minus.5pt \hskip1truemm #\hfil 	
& \hfill# & \hfill# & \hfill#					
& \tabskip=.9pt plus.5pt minus.5pt \hskip1truemm #\hfil 	
& \hfill# 							
& \hfill# 							
& \hfill# 							
& \hfill# 							
& \hfill# 							
& \hfill# 							
& \tabskip=.9pt plus.5pt minus.5pt \hskip1truemm #\hfil 	
& \hfill#\hfill							
& \hfill# \tabskip=0pt \cr					
\noalign{\hrule\smallskip}
& Names 
&& Type 
&& \multispan3 1st component 
&& \multispan3 2nd component 
&& \multispan3 3rd component 
&& \multispan6 Ratios 
&& \multispan1 Central 
&\cr
\noalign{\smallskip}
& 
&& 
&& \hfill $l_1$ \hfill 
& \hfill $\epsilon_1^{\hbox{max}}$ \hfill
& \hfill PA$_1$ 
&& \hfill $l_2$ \hfill 
& \hfill $\epsilon_2^{\hbox{max}}$ \hfill 
& \hfill PA$_2$ \hfill 
&& \hfill $l_3$ \hfill 
& \hfill $\epsilon_3^{\hbox{max}}$ \hfill 
& \hfill PA$_3$ \hfill 
&& \hfill $\beta_{12}$ \hfill 
& \hfill $\theta_{12}$ \hfill 
& \hfill $\gamma_{12}$ \hfill 
& \hfill $\beta_{13}$ \hfill 
& \hfill $\theta_{13}$ \hfill 
& \hfill $\gamma_{13}$ \hfill 
&& \multispan1 J--K 
&\cr
\noalign{\smallskip} 
& \hfill (1) \hfill 
&& \hfill (2) \hfill 
&& \hfill (3) \hfill 
& \hfill (4) \hfill 
& \hfill (5) \hfill 
&& \hfill (3) \hfill 
& \hfill (4) \hfill 
& \hfill (5) \hfill 
&& \hfill (3) \hfill 
& \hfill (4) \hfill 
& \hfill (5) \hfill 
&& \hfill (6) \hfill 
& \hfill (7) \hfill 
& \hfill (8) \hfill 
& \hfill (6) \hfill 
& \hfill (7) \hfill 
& \hfill (8) \hfill
&& \hfill (9) \hfill 
&\cr
\noalign{\smallskip\hrule\smallskip}
& NGC~1097  && B+B   
&& $^I\approx$80 & $^I$0.67 & $^I$147 && 10.3 & 0.46 &  30 &&     &      &     
&&  7.8 &            --63 & $^I$3.8 &    &      &    && no colour
&\cr
& NGC~2681  && B+B   
&& 28.7 & 0.34 &  80 &&  4.9 & 0.11 & 5 &&     &      &     
&&  5.8 &             +75 & 1.8 &    &      &    && o 
&\cr
& NGC~3081  && B+B   
&& 40.9 & 0.65 &  65 && 10.3 & 0.37 & 113 &&     &      &     
&&  4.0 &  {\smallit +48} & 2.3 &    &      &    && rs  
&\cr
& NGC~4314  && B+B   
&& $\approx$75 & 0.69 & 143 && 5.6 & 0.19 & 131 &&     &      &     
&& 13.4 & --12 & 7.5 &    &      &    && r
&\cr
& NGC~5850  && B+B   
&& 83.7 & 0.68 & 112 &&  9.2 & 0.31 &  46 &&     &      &     
&&  9.1 &            --66 & 3.5 &    &      &    && r,rs       
&\cr
& NGC~5905  && B+B   
&& 36.6 & 0.61 &  20 &&  5.7 & 0.19 & 133 &&     &      &     
&&  6.5 &             +67 & 3.8 &    &      &    && $^{H-K}$r,o 
&\cr
& NGC~6782  && B+B   
&& 35.8 & 0.54 & 178 &&  5.2 & 0.40 & 149 &&     &      &     
&&  6.8 &            --29 & 3.0 &    &      &    && $^{J-H}$r
&\cr
 \noalign{\smallskip}
& NGC~470   && B+T
&& 31.6 & 0.55 &  14 &&  7.6 & 0.46 & 150 &&     &      &     
&&  4.1 & {\smallit --46} & 2.9 &    &      &    && rs 
&\cr
& NGC~6951  && B+T   
&& 56.7 & 0.59 & 85 &&  &  &  &&  &  &     
&&  &  &  &  &  &  && r,o 
&\cr
& NGC~7479  && B+T   
&& 45.6 & 0.80 & 6 &&  &  &  &&  &  &     
&&  &  &  &  &  &  && o 
&\cr
 \noalign{\smallskip}
& NGC~2950  && B+T+B 
&& 37.6 & 0.45 & 152 &&        &        &       &&  6.2 & 0.25 &  92 
&&        &         &       &  6.1 &  +60 & 2.4 && - 
&\cr
& NGC~4340  && B+T+B 
&& $^J$51.7 & $^J$0.41 &  $^J$32 &&        &        &       &&  $^J$5.0 & 
$^J$0.10 &  $^J$35 
&&        &         &       & $^J$10.3 &  $^J$--3 & $^J$4.9 && $^{J-H}$bc 
&\cr
& NGC~7098  && B+T+B 
&& 57.3 & 0.57 &  50 &&        &        &       && 14.4 & 0.32 &  71 
&&        &         &       & 3.9 &  --21 & 2.2 && -
&\cr
 \noalign{\smallskip\hrule\smallskip}
}}
\noindent
Column (2): Suggested classification. B = barred structure / T =
twisted structure
\hfill\par\noindent 
Column (3): Projected semi-major axis of the component $i$ in arcsec
defined at the minimum of ellipticity.  Values coming from other
bands are preceded by the corresponding letter
\hfill\par\noindent
Column (4): Projected maximum ellipticity of the component $i$
\hfill\par\noindent
Column (5): Projected position-angle of the component $i$ in degree at
its maximum of ellipticity 
\hfill\par\noindent
Column (6): Projected semi-major axis length ratio of the
component $i$ to the component $j$ 
\hfill\par\noindent
Column (7): Projected relative angle between the component $i$ and the
component $j$. A positive angle means the structure $j$ is ``leading''
the structure $i$ while a negative angle mean it is``trailing''.
Values in italic indicate that the sense of rotation is unknown and
the trigonometric convention is used
\hfill\par\noindent
Column (8): Integrated luminosity ratio of the component $i$ to the
component $j$ 
\hfill\par\noindent 
Column (9): bc = blue center / o = other features / r = red ring / rs
= red secondary bar
\hfill\par\noindent }

\endtab 
 
\titled{NGC~2950.} 
The 0.9\arcsec\ resolution of our images prevents an accurate
determination of the structural parameters of the secondary bar.  The
signature of this small bar appears less pronounced in K-band than in
I-band. The profiles are thus more noisy. Moreover, the image taken in
the H-band is saturated. However, in the region of the expected
secondary bar, the \PA\ really shows a plateau, especially pronounced
in the J-band.  The J--K colour map shows an unusual structure. The
red emission is indeed aligned with the primary bar although its
distribution is asymmetric. There is no red emission towards the south
part of the primary bar.
 
\titled{NGC~3081.} 
In the K-band, the secondary bar observed in optical bands by Buta
(1990) as well as in Paper I is clearly confirmed.  It is also very
luminous with respect to the primary bar. Indeed, it creates a very
significant bump in the surface brightness profile.  The nuclear ring
does not appear in the K-band, suggesting that it is not luminous
enough and it is better traced by dust and/or star formation than by
old stars.  The secondary bar also appears in the J--K colour map as a
very red and elongated structure (J--K~$\approx 1.0$).  Note that the
H-image is not photometric.
 
\titled{NGC~4314.} 
Using the HST telescope, Benedict \etal\ (1993) have found an oval
distortion of 4\arcsec\ (semi-major axis length) in the nuclear
region. However, with our JHK data at 0.9\arcsec\ resolution it is
impossible to confirm the presence of such an oval. We find a more or
less boxy shape of the inner isophotes due to the spots of star
formation along the nuclear ring. We cannot resolve these spots
because of the blurring by seeing.  The major-axis of these boxy
isophotes is aligned with the primary bar.  This creates the small
innermost maximum on the ellipticity profiles at 4.8\arcsec\ with
roughly the same \PA\ as the primary bar ($\theta\!=\!-12\degr$).
This innermost ellipticity maximum depends on the wavelength and could
therefore be biased by dust absorption.  We have nevertheless decided
to associate the innermost structure with the secondary bar of
Benedict \etal\ (1993) because of their 0.13\arcsec\ resolution,
although it is not clear why this object shows the largest values of
$\beta$ and $\gamma$ (cf. Table~\tabresult).  Moreover, on HST images
$\theta\!=\!-3\degr$.  The nuclear ring is clearly visible in the J--K
colour map ($0.25 \la \rm J-K \la 0.40$).  Shaw \etal\ (1995) claimed
that the nucleus of this object is blue inside 2\arcsec.  Due to our
point spread function we are unable to confirm their result. The red
nucleus visible in our J--K colour map could be an artifact: all stars
in the field show similar features.
 
\titled{NGC~4340.} 
The innermost structure regarded as a secondary bar in Paper I seems
to be also present in JH-images (K-band is saturated). It is roughly
aligned in projection with the primary bar ($\theta_{13} \approx
-3\degr$).  The J--H colour map displays a blue central region
(J--H~$\approx -0.1$) as compared with the rest of the primary bar
(J--H~$\approx 0.1$).
 
\titled{NGC~5850.} 
Without any surprise, the JHK-images confirm the double-barred nature
of this galaxy.  In particular, the secondary bar appears in the J--K
colour map as a redder structure as well.  It is also surrounded by a
very faint red ring which is difficult to see on the greyscale figure.
 
\titled{NGC~5905.} 
The secondary bar of NGC~5905 is confirmed on H- and K-images with the
same value for the maximum ellipticity (no J-image). At the end of the
secondary bar a very red circumnuclear ring appears in the H--K colour
map. This ring is correlated with the H$\alpha$ and the blue B--I ring
observed in Paper I.  However, the reddest part of the H--K ring
(H--K~$\approx 0.8$) does not coincide with the bluest regions of the
B--I ring (B--I~$\approx 1.8$) contrary to most galaxies with nuclear
rings. NGC~5905 is one of our two cases with such phenomenon
(cf. NGC~6951).
 
\titled{NGC~6782.} 
This object clearly is a double-barred galaxy. Indeed we detect the
secondary bar surrounded by a stellar ring in all three bands.  The
dust distribution seen in B--I colour map (Paper I) is thus better
explained by taking into account this bar+ring structure.  Between the
two bars, the ring creates a bump in the \PA\ profiles. It does not
have the same projected orientation as the secondary bar.  The squares
appearing on the images result from the bad removal of bright
foreground stars on sky-frames.  The plateau in the outermost part of
the surface brightness profile in the K-band is due to a change of the
weather conditions. The sky brightness is thus poorly subtracted. We
have computed the J--H colour map instead of the J--K one. The nuclear
ring (J--H~$\approx 0.5$) is clearly visible as a redder structure
than the primary bar (J--H~$\approx 0.3$).
 
\titled{NGC~6951.} 
We cannot unambiguously confirm the secondary bar suggested in Paper I
due to the strong isophote twists.  This twisted structure is
surrounded by a red (J--K~$\approx 1.35$) nuclear ring visible on J--K
colour map while the surrounding region has J--K~$\approx 1.0$.  The
morphology of this ring is roughly the same on H$\alpha$, B--I and
J--K maps. Moreover, the VLA radio maps of Saikia \etal\ (1994) show a
spiral-like structure in the gas component which is compatible with
the presence of a bar.  However, the biggest J--K spot located N (on
the original J--K map but badly reproduced on the greyscale figure) is
not exactly coincident with the corresponding one in the B--I map
(cf. NGC~5905).  Also, the reddest part of the J--K ring is in the SW,
where the dust darkening is the greatest.  The various hotspots are in
fact spectacularly resolved into numerous smaller star forming regions
with HST V-image (Barth \etal\ 1995).  Their image does not show any
secondary bar.  It is thus not clear if the observed twists could
result from the few intense sites of star formation along the nuclear
ring and/or the dust pattern inside the ring.
 
\titled{NGC~7098.} 
The B+T+B classification of Paper I is confirmed. Both maxima in
ellipticity are well separated by a triaxial bulge. This bulge shows a
twist of almost 25\degr. The J--K colour map does not display any
peculiar structure.
 
\titled{NGC~7479.} 
It is not possible to detect reliably any structure apart from the
main bar.  This galaxy is already too inclined (see Sect.~4.2) and the
equatorial dust obscures the region where the secondary bar detection
could be possible.  A careful inspection of the J--K colour map
indicates that the whole galaxy is crossed by a dust lane. However,
the spatial distribution seems not to be symmetric and the central
part shows a clear isophote twists.
 
\titleb{Global properties} 
The various projected structural parameters in the K-band for the
galaxies discussed in Sect.~3.1 are summarized in Table~{\tabresult}
and displayed in Figs.~{\FBETH}, {\FBEGA} and {\FEPSI}.
 
For the category {B+B} including NGC~470, $4.0 \leq \beta \leq 13.4$
and $1.8 \leq \gamma \leq 7.5$, and a typical double-barred galaxy has
$\beta \approx 7.2$, $\gamma \approx 3.5$, $e_s^{\rm max} \approx
0.31$, and $e_p^{\rm max} \approx 0.58$.  For the category {B+T+B},
$3.9 \leq \beta_{13} \leq 10.3$ and $2.2 \leq \gamma_{13} \leq 4.9$,
and mean values $\beta_{13} \approx 6.8$, $\gamma_{13} \approx 3.2$,
$e_3^{\rm max} \approx 0.22$, and $e_1^{\rm max} \approx 0.48$.  As
already observed for the I-band (Paper I), generally $e_s^{\rm max}
\!<\!e_p^{\rm max}$. This may be an observational effect due to 
seeing and pixel size which affect most central structures. But, on 
the other hand galactic centers are dynamically hot and the formation 
of weaker bars are not unexpected in that case (see e.g. Athanassoula 
1983). 
 
In their sample, Shaw \etal\ (1995) have found a majority of galaxies
with blue nuclear regions in the near-IR colours J--H and H--K.  On
the contrary, this feature is not dominant in the sample presented
here since only 1 galaxy (NGC~4340) in 13 has a blue centre in J--K.
NGC~4340 has also a blue centre in J--H. Thus, its colour does not
result from the central saturation of the K-image. Moreover, the blue
region is larger than the saturated one. Another galaxy (NGC~2681) has
a blue centre of a few pixels but this is surely due to the Seyfert
nucleus (cf. Paper I).  On the contrary, a red secondary bar is
clearly observed in 3 galaxies, and a red nuclear ring in 5 galaxies.
 
\begfig 8.8cm 
\nomps{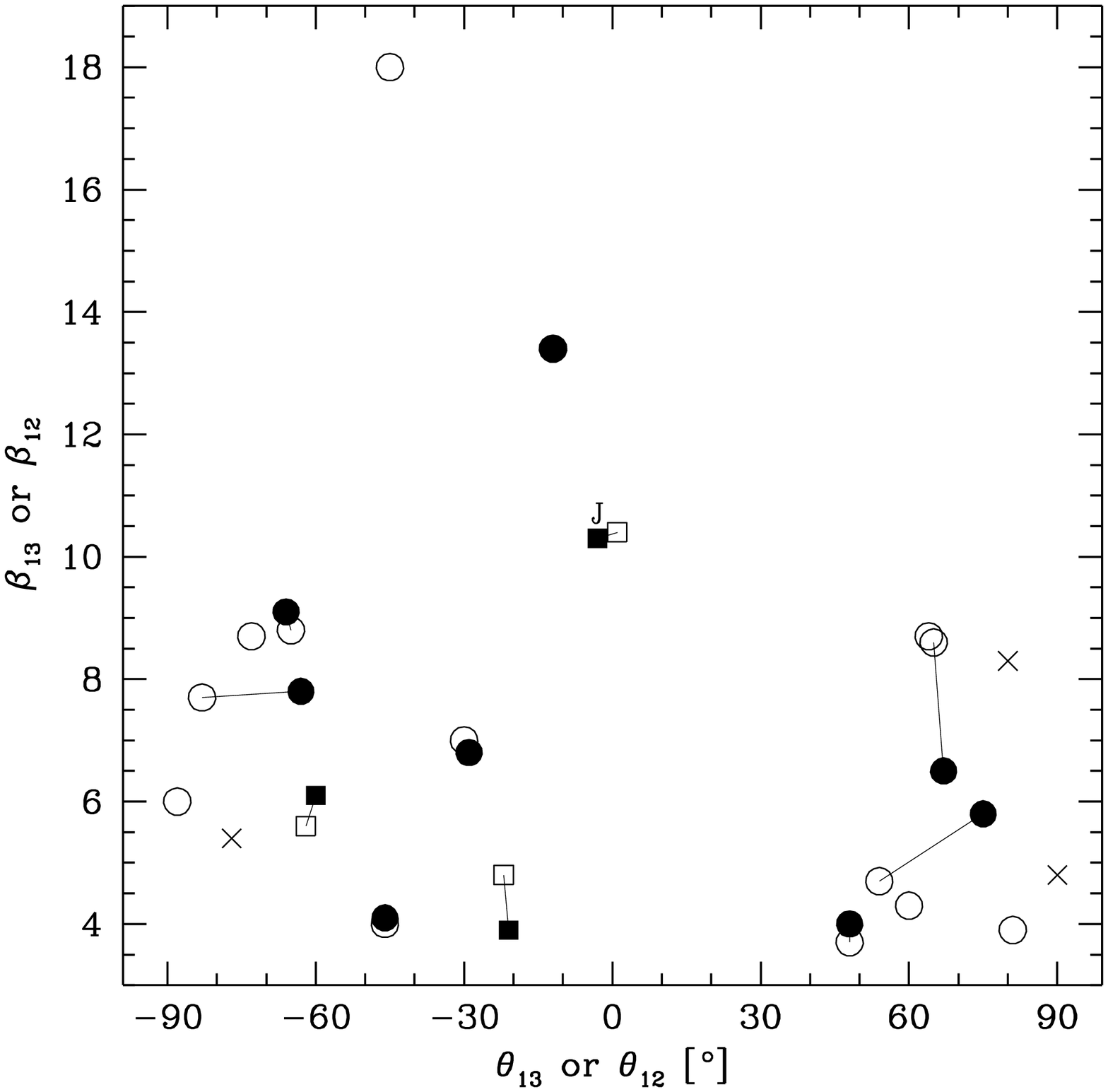} 
\figure{\FBETH}{ 
Relation in the K-band between $\theta_{12}$ and $\beta_{12}$ (full 
circles) and $\theta_{13}$ and $\beta_{13}$ (full squares). The open 
circles and squares correspond to the values found in the I-band 
(Paper I).  The solid lines link identical galaxies.  The crosses 
correspond to various numerical simulations.  The J letter indicates a 
measurement in the J-band } 
\endfig 
 
\begfig 8.8cm 
\nomps{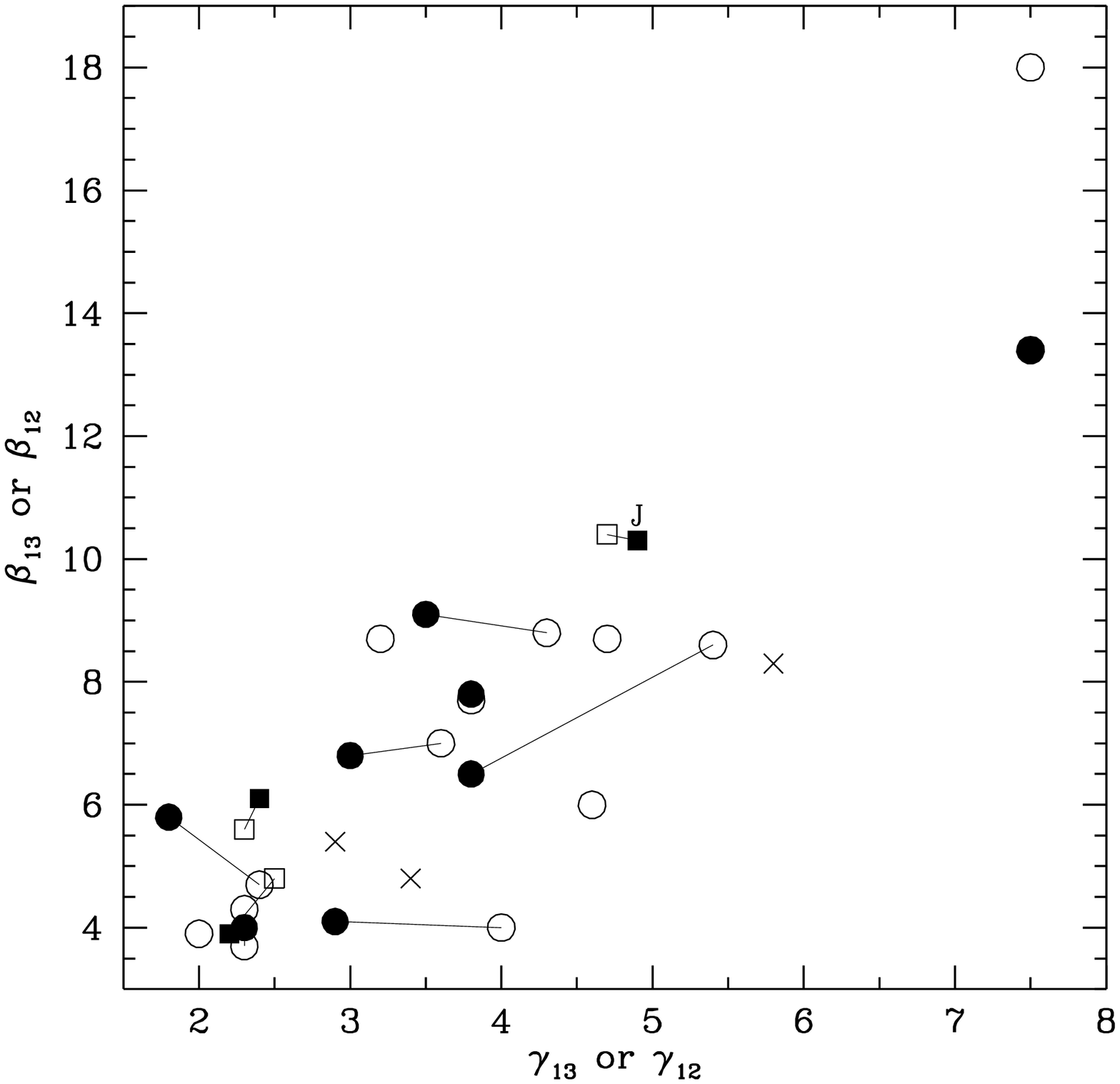} 
\figure{\FBEGA}{ 
The same as Fig.~{\FBETH} but for the relation between $\gamma_{12}$ 
and $\beta_{12}$ (full circles) and $\gamma_{13}$ and $\beta_{13}$ 
(full squares) } 
\endfig 
 
\begfig 8.8cm 
\nomps{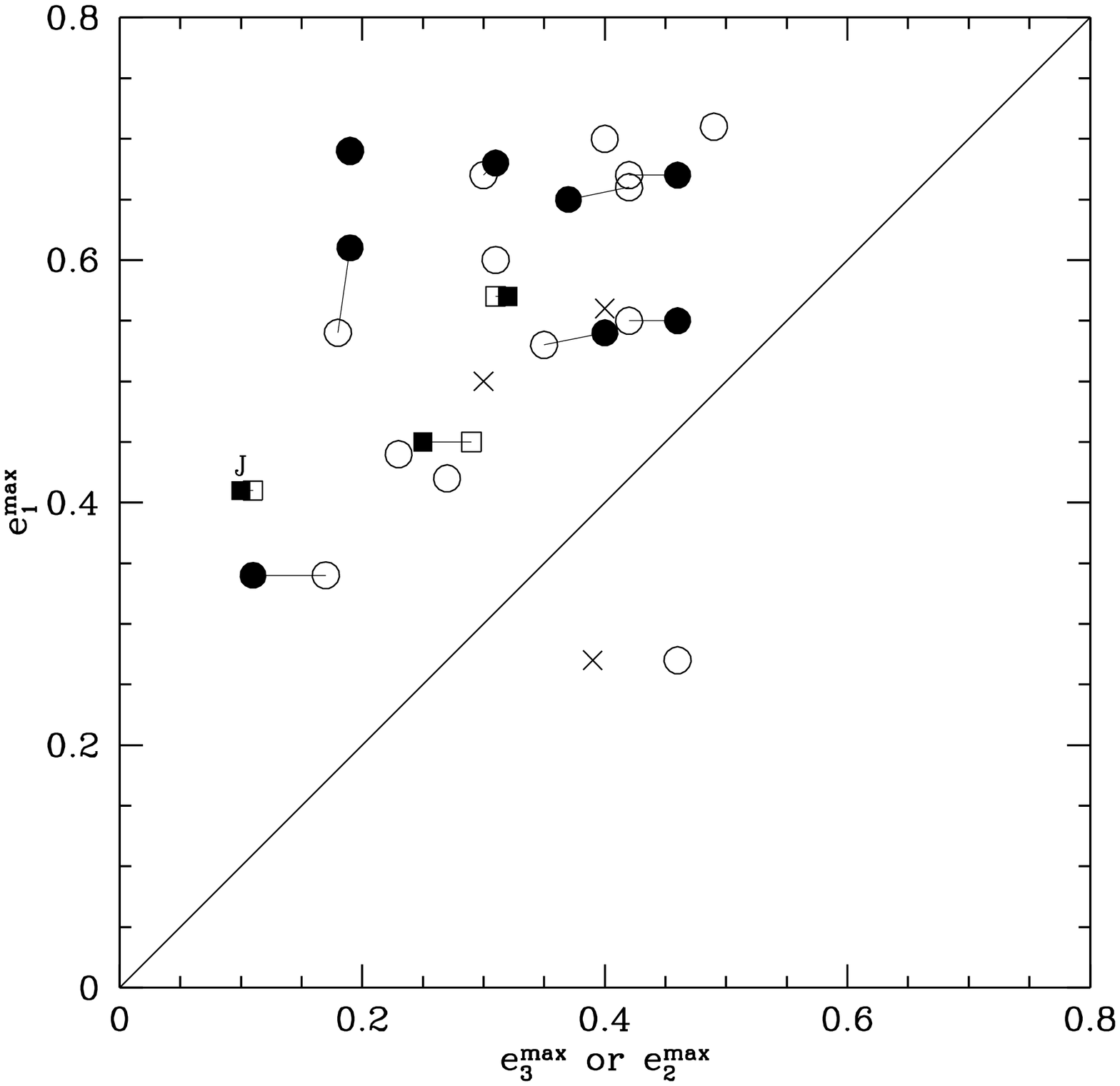} 
\figure{\FEPSI}{ 
The same as Fig.~{\FBETH} but for the relation between $e_1^{\rm max}$ 
and $e_2^{\rm max}$ (full circles) and $e_1^{\rm max}$ and $e_3^{\rm 
max}$ (full squares) } 
\endfig 
 
\begfig 12cm 
\nomps{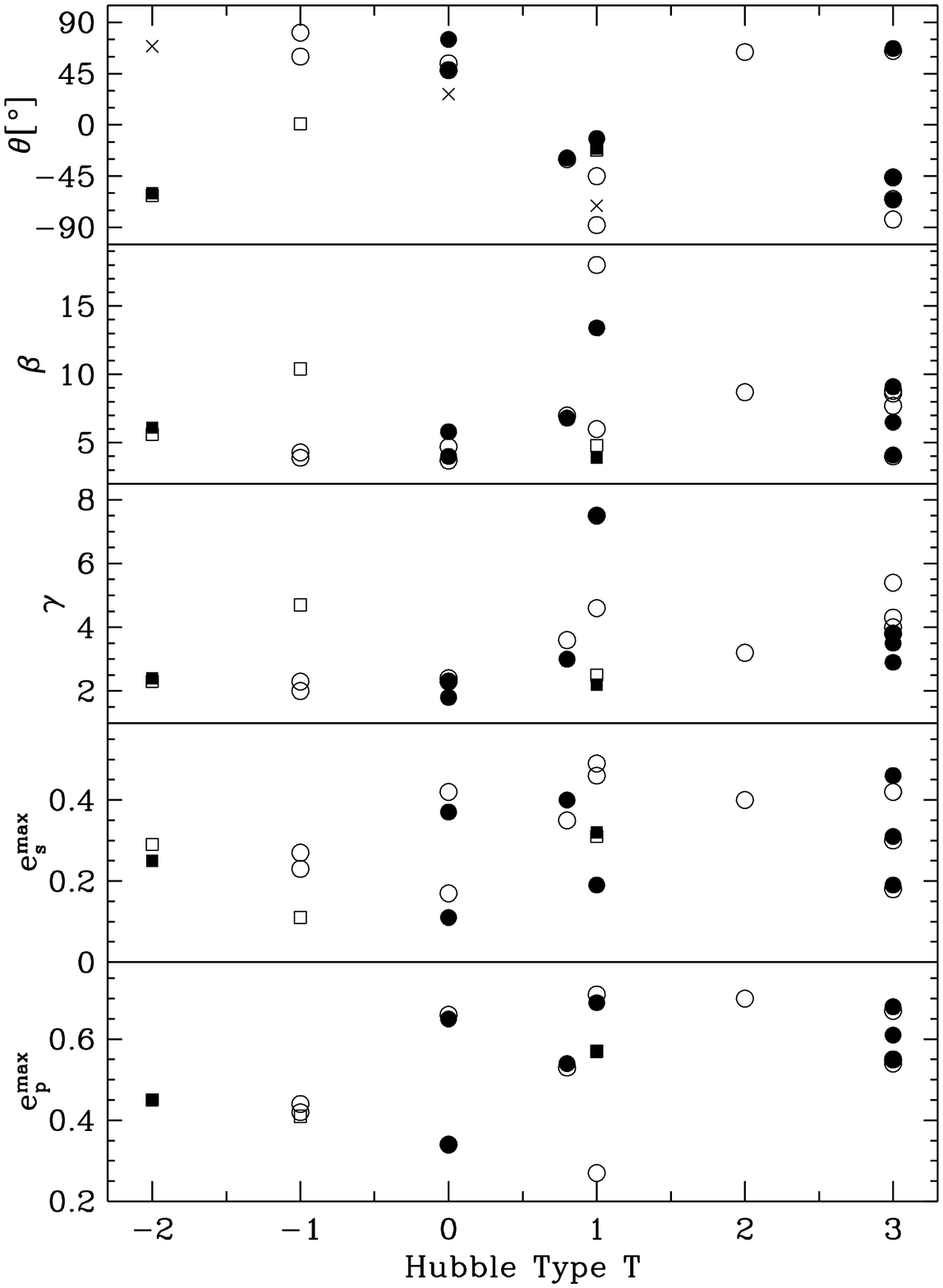} 
\figure{\FTYPE}{ 
Projected values of $\theta$, $\beta$, $\gamma$, $e_s^{\rm max}$, and 
$e_p^{\rm max}$ as a function of the Hubble type T. The full circles 
are for the K-band whereas the open circles correspond to the I-band 
(Paper I).  The crosses come from Buta \& Crocker (1993) } 
\endfig 

Due to the poor and biased statistics, no reliable percentage of
double-barred galaxies among barred galaxies can be given. In recent
years the number of such galaxies has strongly increased due to the
advent of optical CCD's and infrared arrays. This certainly indicates
that this feature might be more common than previously estimated.
Figure~{\FTYPE} shows the dependence of the projected structural
parameters with respect to Hubble type T of the galaxies both in the
I- and K-bands. In our sample, no double-barred galaxies have been
found with types later than T=3.  No preferential angle $\theta$ has
been found and there is no correlation between the Hubble type and
$\theta$.  In particular, leading secondary bars are not preferably
found in late-type spirals. In the I-band sample of 13 double-barred
galaxies, 6 were leading and 7 trailing. The above results do not
change after deprojection (see Sect.~3.4).  With a few exceptions and
a relatively high scatter, early-type galaxies seem to have lower
$\beta$ and $\gamma$ than late-type galaxies.  The best correlation is
observed for $\gamma$ in the K-band but the number of galaxies is
still small. This can easily be explained by the fact that the Hubble
sequence is also a sequence of bulge to disc luminosity (or mass)
ratio (increasing bulge to disc luminosity with decreasing T). The
maximum ellipticity of the secondary bars does not seem to correlate
with the Hubble type whereas late-type galaxies could have higher
maximum ellipticity.
 
\titleb{Comparisons with BVRI} 
Although the values of the structural parameters $\theta$ and $\beta$
are noticeably the same between Paper I (I-band) and the present paper
(K-band), a few pronounced discrepancies deserve an
explanation. Indeed, for NGC~1097 we find a difference of 20\degr\ for
$\theta$, while $\beta$ increases from 7.7 to 7.8. This could be
explained by the presence of dust which leads to an erroneous
determination in the I-band as the darkening alters the morphology of
the isophotes.  For NGC~5905, the same explanation holds but the dust
essentially modifies the $\beta$ ratio (6.5 instead of 8.6 in I-band),
as $\theta$ agrees between I- and K-bands within 5\degr. Note that in
both cases, the primary and the secondary bars are longer in K-band
than in I-band. Again, dust darkening could explain the differences
both in $\beta$ and $\theta$ for NGC~2681 (for $\theta$, 75\degr\
instead of 54\degr\ in I-band; for $\beta$, 5.8 instead of 4.7 in
I-band). The dust distribution does not follow regular lanes as in
most other barred galaxies but is rather distributed on fine thread
over the whole disc. The dust is also visible in the K-band.
 
The observed intervals of $\beta$ and $\gamma$ are very similar in I-
or K-bands with a distinct upper limit given only by one galaxy, i.e.
NGC~5728 (I-band) and NGC~4314 (K-band). These galaxies are either
very peculiar, or there is an observational bias toward the lowest,
easiest to detect, $\beta$ values.  Globally, a typical (mean)
double-barred galaxy has similar values for its structural parameters
in I- or K-bands.  However, in the subsample measured in I- and
K-bands, the majority of $\gamma$ are smaller in the K-band than in
the I-band (see Fig.~{\FBEGA}) with respective mean values of 2.9 and
3.5.  Since, the K-band is a better tracer of the stellar mass
distribution, one is led to conclude that the central mass
concentration is slightly underestimated in the I-band.  Linear fits
show that the slope of the relation between $\gamma$ and $\beta$ is
similar for both bands.
 
The mean values for secondary and primary bar ellipticities are about
10\% lower in K-band than in I-band.  As far as we are concerned with
the ellipticity of the secondary bar, the differences between I- and
K-bands could arise from the different seeing conditions and
resolutions (that give different seeing samplings). For ESO images,
taken at 0.49\arcsec\ resolution and a seeing of roughly 1\arcsec, all
four galaxies have a greater maximum ellipticity for the secondary bar
in the K-band. This is not the case for the images taken at
0.6\arcsec\ and 0.9\arcsec\ of resolution with 1.5\arcsec --
1.7\arcsec\ of seeing which show a decrease of $e_s^{\rm max}$.
 
\begfig 8.8cm 
\nomps{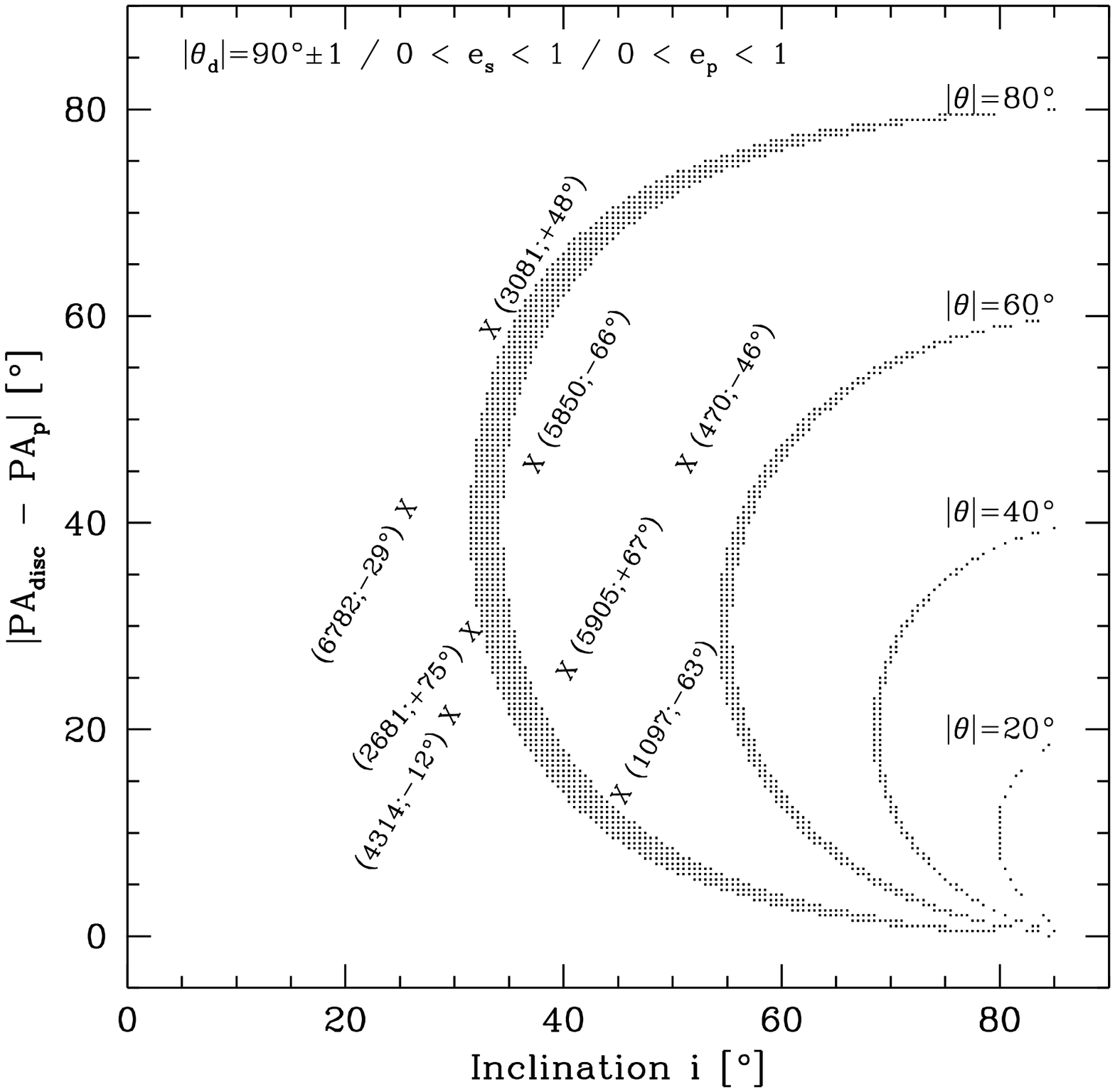} 
\figure{\FDEPRO}{  
The regions where the deprojected angle between the two bars 
$|\theta_d|\!=\!90\degr$ are displayed in the diagram $|\PAm_{\rm 
disc} - \PAm_p|$ versus $i$ for various observed angle $\theta$ and 
any ellipticities.  Real galaxies are plotted with a cross and a label 
(NGC~number; $\theta$). No one is close to these regions } 
\endfig 

Finally, it should be noted that for the category {B+B} primary bar
lengths measured in the K-band are greater than the ones in the
I-band. The amplitude of this effect is in no case greater than about
13\%. This result has been carefully checked since we have scaled the
K-band pixel size (expressed in $(\arcsec) \cdot \rm px^{-1}$) with
respect to the value in the I-band. Assuming that the I-band pixel
sizes (one for each telescope used) are free of errors, we have
measured positions of at least 3 stars visible in both I- and
K-images. The geometric transformation from I-band coordinates to
K-band ones give us the true pixel size. We systematically found
significant differences with the theoretical value or the value given
by observer's manuals. The lengths given in Table~\tabresult\ use the
true pixel size.
 
\titleb{Deprojections} 
How to deproject images is a long-standing problem we do not want to
develop here (see e.g. Binney \& Gerhard 1995, and references
therein).  However, it is necessary to check that the deprojected
angles between the two bars $\theta_d$ really do not take any peculiar
value, in particular 90\degr.  Deprojections are performed as in Paper
I.  Figure~{\FDEPRO} displays, in the diagram $|\PAm_{\rm disc} -
\PAm_p|$ versus the inclination angle $i$, the regions where the
deprojected angles between the two bars $|\theta_d|\!=\!90\degr$ for
various observed angle $\theta$ and any ellipticities.  Some of the
observed galaxies are also plotted and one can see that none is close
to these regions. The two bars are really misaligned. After
deprojection, there is still roughly half leading and half trailing
secondary bars. This is a crucial point for the understanding of the
bar-within-bar phenomenon (see Sect.~5).
 
\titlea{Numerical simulations}
 
Thanks to numerical simulations including stars and gas, Friedli \&
Martinet (1992, 1993; Combes 1994) have shown that the secondary
stellar bar can be created from a decoupling of the central dynamics
triggered by gas flood accreted towards the centre by the primary
stellar bar. The two bars have then two different pattern speeds with
the secondary bar being the faster.

We have computed a new set of simulations which mainly differs from
the one by Friedli \& Martinet (1993) in the following way: 1) Model
$B_{\rm no}$. The particle number is doubled, i.e. $N_g\!=\!20~000$,
$N_*\!=\!200~000$, and the mass of the various components has been
decreased, i.e. $M_g\!=\!0.055$ (instead of 0.1), $M_b\!=\!0.05$
(0.1), $M_d\!=\!0.50$ (0.8). 2) Model $B_{\rm sf}$. The same as model
$B_{\rm no}$ but star formation is included (see Friedli \& Benz 1995
for details). We analyse below in the same way as the observations
(i.e. with the ellipse fitting technique) various 2D density
projections of the 3D numerical simulations.  Only star particles have
been used unless otherwise stated. As for the observations, the 2D
grid used is 256$\times$256.  Of particular interest are the time
evolution and orientation changes of the structural parameters.  The
typical evolution of the stellar isodensity contours can be found in
Fig.~[4] of Friedli \& Martinet (1993).

\begfig 12cm 
\nomps{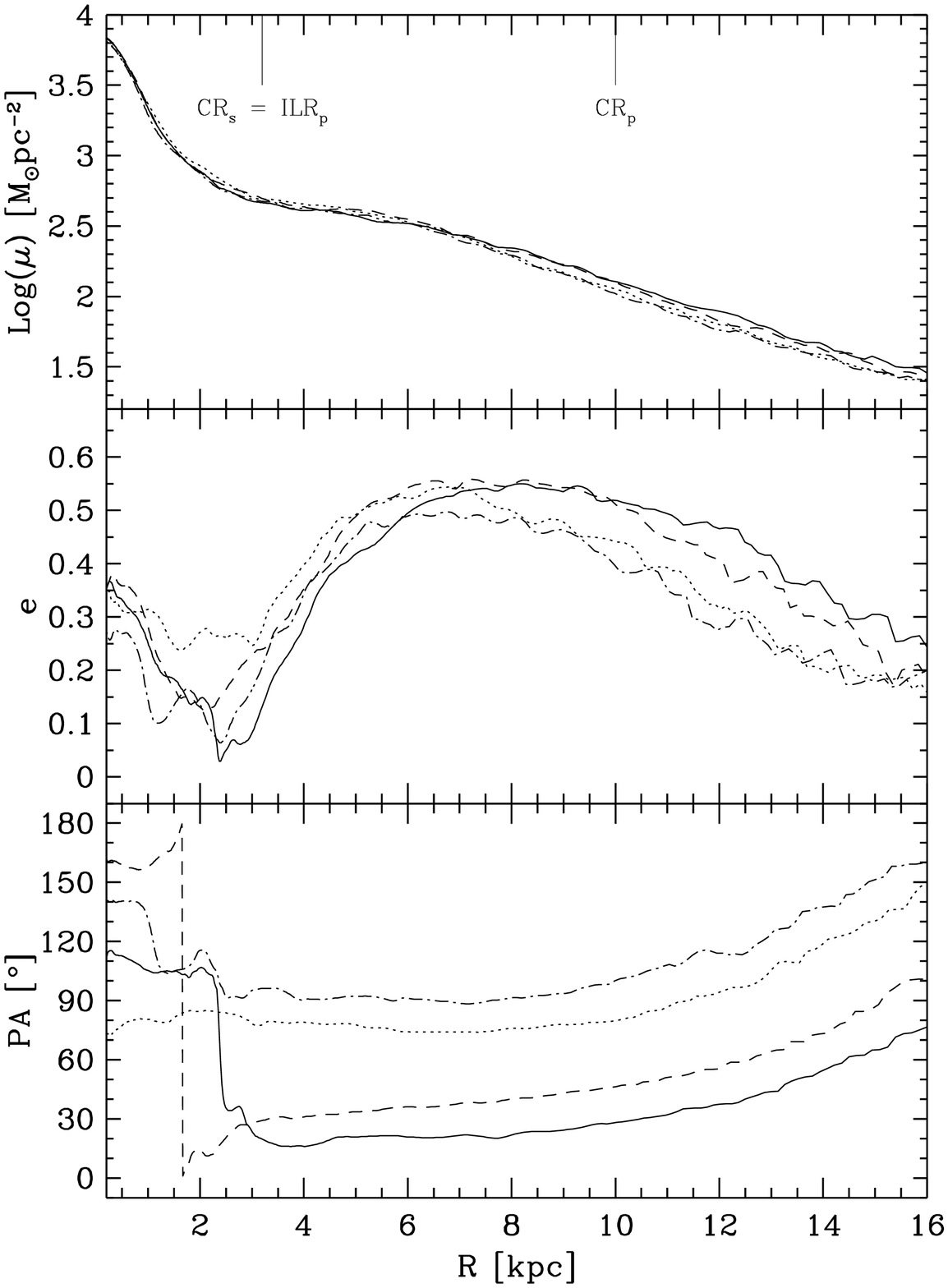} 
\figure{\FNSTIM}{ 
Plots of stellar surface density $\mu$, ellipticity $e$ and
position-angle \PA\ profiles as a function of the semi-major axis of
the fitted ellipse, at various times $t\!=\!1270$~Myr (solid curve;
$\theta \approx 90\degr$), $t\!=\!1225$~Myr (dotted curve;
$\theta \approx 0\degr$), $t\!=\!1255$~Myr (dashed curve; $\theta
\approx 60\degr$), and $t\!=\!1210$~Myr (dot-dashed curve; $\theta
\approx -50\degr$).  Model $B_{\rm no}$ seen face-on }
\endfig 
 
\begfig 12cm 
\nomps{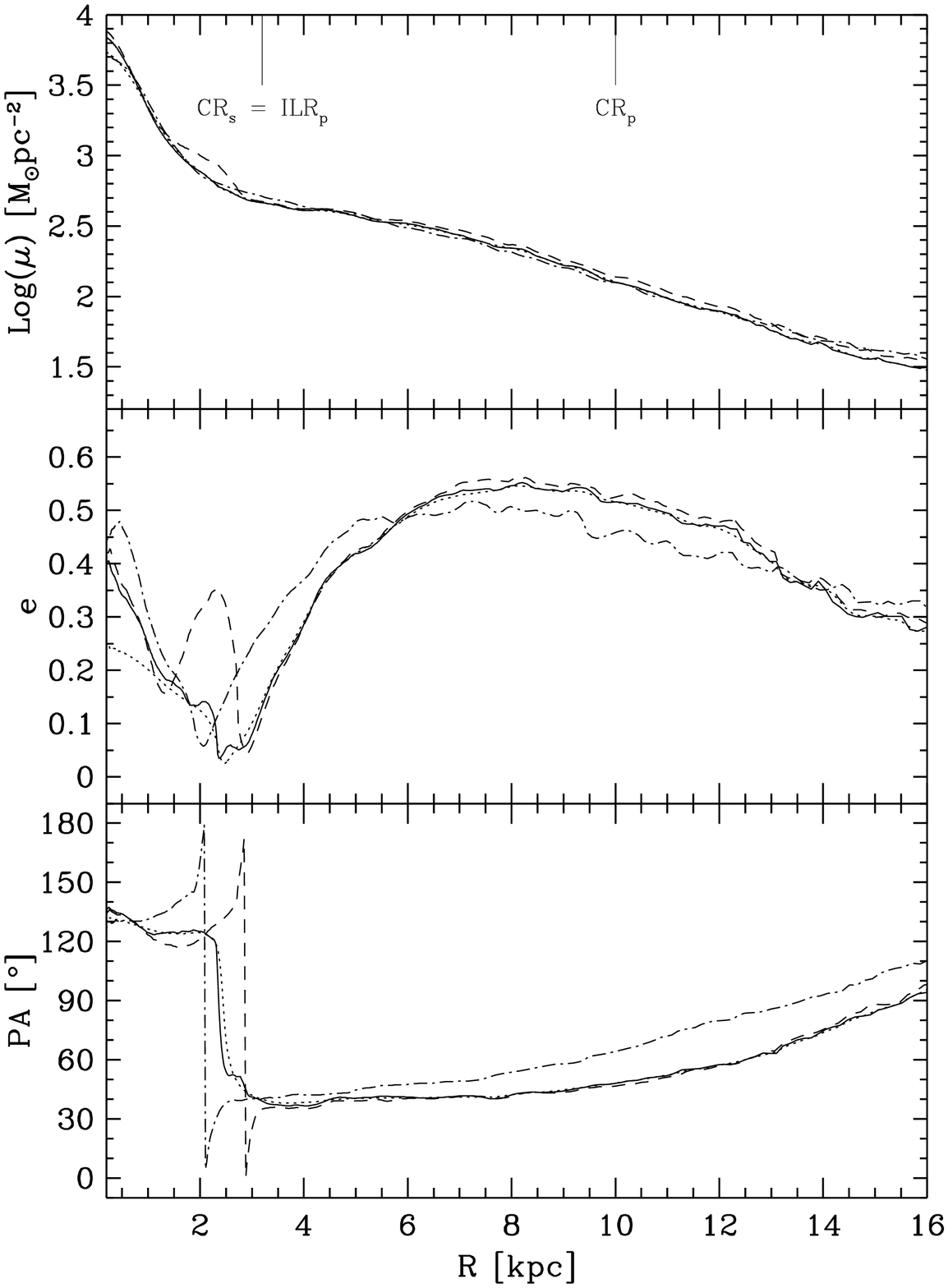} 
\figure{\FNSCOM}{ 
Comparison of $\mu$, $e$ and \PA\ profiles as a function of the
semi-major axis of the fitted ellipse between the model $B_{\rm no}$
($t\!=\!1270$~Myr) with low smoothing (solid curve), with high
smoothing (dotted curve), with gas mass included (dashed curve), and
the model $B_{\rm sf}$ ($t\!=\!1390$~Myr; dot-dashed curve). Galaxies
are seen face-on }
\endfig 
 
\begfig 12cm 
\nomps{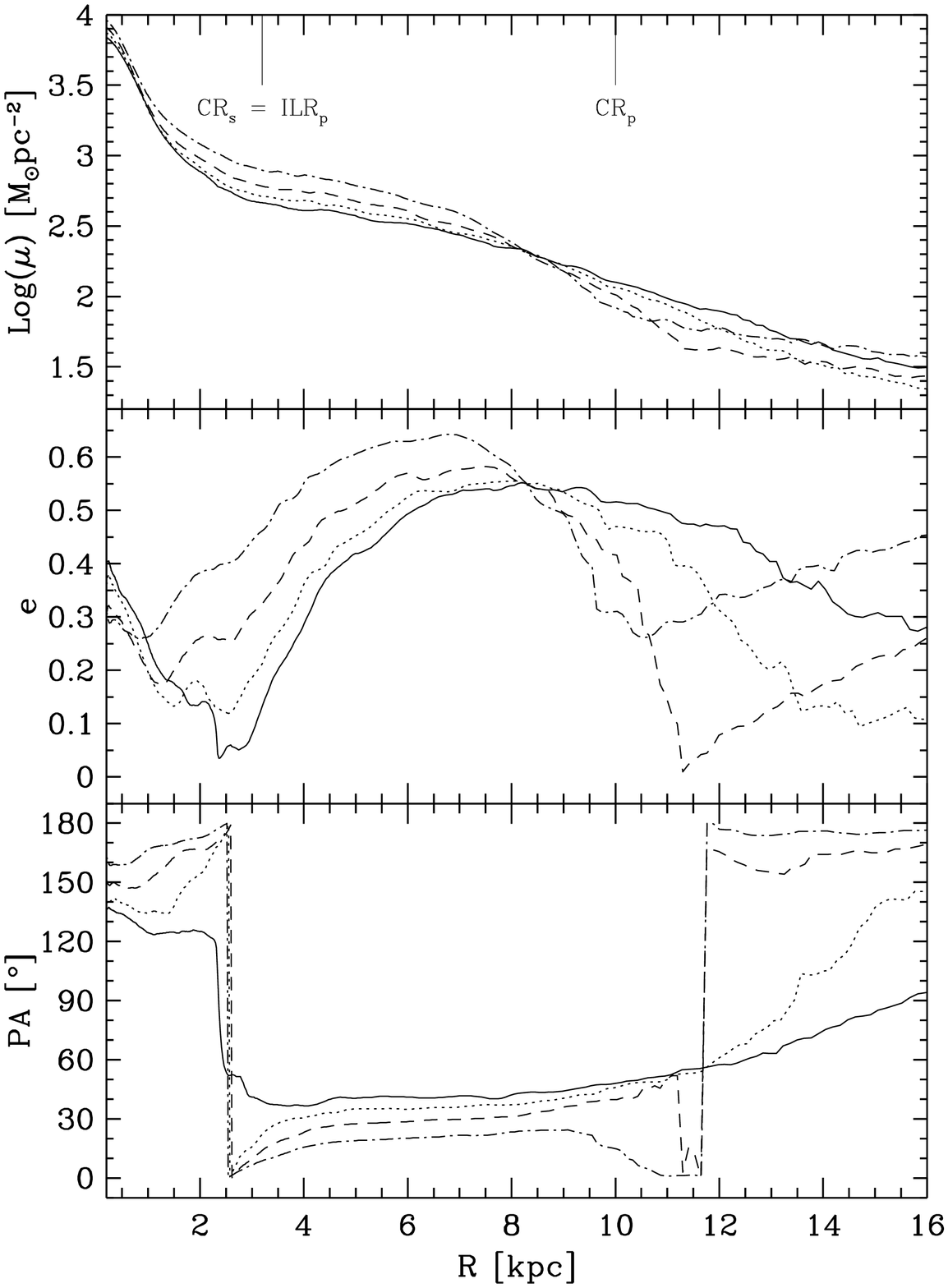} 
\figure{\FNSORI}{ 
The same as Fig.~{\FNSTIM} but at various inclinations $i\!=\!0\degr$
(solid curve), $i\!=\!30\degr$ (dotted curve), $i\!=\!45\degr$ (dashed
curve), and $i\!=\!60\degr$ (dot-dashed curve). Model $B_{\rm no}$ at
$t\!=\!1270$~Myr }
\endfig 

\titleb{Generic model} 
The time evolution of the structural parameters of our generic model
$B_{\rm no}$ is presented in Fig.~{\FNSTIM}.  The time dependence of
$\theta$ is clearly seen and indicates the presence of two different
pattern speeds.  At first order, these numerical models fairly well
reproduce the observed features of double-barred galaxies (two
distinct PA's, maximum ellipticities, and surface brightness
slopes). The values of $\beta \approx 4.8$, $\gamma \approx 3.4$,
$e_s^{\rm max} \approx 0.40$ and $e_p^{\rm max} \approx 0.56$ are also
very similar to what is observed (see Figs.~{\FBETH}, {\FBEGA},
{\FEPSI}).

Let us concentrate on two peculiar angles: 1) $\theta \approx
90\degr$.  This particular orientation leads to the best
determination of the end of the secondary bar by looking at the
position of the distinct minimum ellipticity (close to zero).  The
transition from $\PAm_s$ to $\PAm_p$ is also very sharp.  2) $\theta
\approx 0\degr$. By definition, there is no transition from $\PAm_s$
to $\PAm_p$. The ellipticity dip between the two maxima is strongly
reduced. Thus, it could be very difficult to highlight galaxies with
nearly parallel bars within bars, even in the face-on ones, with
morphological criteria only.

Contrary to the end of the secondary bar, the end of the primary bar
is generally difficult to infer from ellipse fits mainly because of
the significant amplitude of the spiral arms (the ellipticity does not
go to zero and the \PA\ is progressively twisted).  So, we chose to
associate the primary bar end with its corotation radius.  Another
problem is that near the primary bar end both the grid resolution and
the particle density are becoming poor.  It is then necessary to
smooth slightly the surface density which can alter the values of the
structural parameters, in particular the secondary bar ellipticity
(cf. Fig.~{\FNSCOM}). This smoothing effect is similar to the one
induced by seeing or pixel size in real observations.

Between the two bars, a massive non-circular broad gaseous ring is
present (see Fig.~[3] of Friedli 1996). It is located inside the
CR$_s$ and is in fact essentially controlled by the secondary bar,
i.e. it has the same pattern speed and orientation. Its contribution
to the total surface density is not negligible as can be seen on
Fig.~{\FNSCOM} where ring signatures are clearly visible: A new
maximum appears in the ellipticity profile ($e_{\rm ring}^{\rm max}
\approx 0.35$) as well as significant twists in the \PA\ profile.
Some gas-induced distortions in the stellar isophotes are also
observed.

\titleb{Projection effects} 
One of the big advantages of 3D numerical models is the possibility to
perform 2D projections in whatever direction.  Figure~{\FNSORI} shows
the changes that are induced in the structural parameters by the
progressive increase of the galaxy inclination $i$ for the model
$B_{\rm no}$ at $t\!=\!1270$~Myr and with $|\PAm_{\rm disc} - \PAm_p|
\approx 45\degr$. When $i$ increases, the \PA\ of the secondary bar
becomes more and more twisted, $e_s^{\rm max}$ decreases and the
ellipticity is no longer close to zero between the two bars (the
ellipticity dip nearly vanishes).  These twists clearly indicate that
the secondary bar is a 3D triaxial structure, and not simply a 2D thin
bar. The most striking effect is the disappearance of distinct
double-bar characteristics as soon as $i \ga 45\degr$, although
$\theta \approx 90\degr$ in this case which is a very favourable
situation.  It is thus very difficult to highlight bars within bars in
such galaxies at least with morphological criteria only.  Moreover,
the maximum inclination allowed for an easy detection of double-barred
galaxies depends both on $\theta$ and $e_s^{\rm max}$ whose values
should not be too small.
 
\titleb{Role of star formation} 
The main differences between the generic model $B_{\rm no}$ (at
$t\!=\!1270$~Myr) and the one with star formation $B_{\rm sf}$ (at
$t\!=\!1390$~Myr) can be seen on Fig.~{\FNSCOM}. The time evolution of
the two models being slightly different, the comparison cannot be
synchronized.  Star formation mainly occurs along spiral arms outside
the primary bar corotation as well as along the secondary bar major
axis as well as along the nuclear ring inside the secondary bar
corotation.  As a consequence, these numerous new stars induce a
significant increase of $e_s^{\rm max}$.  These new stars are
dynamically cold and form a much thinner bar than the one made of the
initial, dynamically hot, old stars.  The secondary bar is also
shorter whereas $e_p^{\rm max}$ is slightly decreased.  Another change
concerns the significant decrease of the gas influence onto stellar
isophotes since about half of its mass has been turned into stars
inside the CR$_s$.
 
\titlea{Discussion} 
The general problem of misaligned, aligned, or twisted structures
within primary stellar bars have been partially reviewed both
observationally and theoretically (Friedli \& Martinet 1993; Friedli
1996; and references therein). In this discussion, the various
possibilities are synthesized in view of the results of this paper.
 
\titleb{Misaligned secondary bars} 
Many explanations can be put forward in order to explain the
phenomenon of secondary misaligned stellar bars (i.e. $\theta_d \neq
0\degr$) within primary stellar bars.
 
\noindent 
1) The observed misalignment is due to projection effects: 
 
\noindent 
-- There are only one primary bar and a round centre which simply 
appeared barred in projection.  In the majority of cases, this 
possibility does not hold since the position angle of the secondary 
structure clearly differs from that of the disc. 
 
\noindent 
-- The two bars are simply perpendicular (i.e. $\theta_d \!=\!
90\degr$) such that the primary bar is made of orbits trapped by the
$x_1$ family, and the secondary bar is built with orbits trapped by
the $x_2$ family.  This can be true in some specific cases
(e.g. NGC~1317) but numerous examples of nearly face-on galaxies with
small $\theta$ (e.g. NGC~1291, NGC~6782; see also Fig.~{\FDEPRO}) rule
out this possibility as a universal explanation. In general, $\theta_d
\neq 90\degr$.
 
\noindent 
2) This peculiar morphology is an observed artifact due to specific 
central patterns of star formation or dust absorption. This argument 
is generally to be excluded since, in the majority of cases, this 
phenomenon is observed in the near-IR as well. 
 
\noindent 
3) A permanent misalignment of two purely stellar bars with the same 
pattern speed is very unlikely due to the very short lifetime 
expected. Strong gravitational torques will indeed quickly align the 
two bars. 
 
\noindent 
4) Shaw \etal\ (1993; see also Combes 1994) have suggested that the
gravitational potential generated by large amounts of leading (with
regard to the primary bar) gas around the ILR of the galaxy could be
responsible of the presence of secondary or twisted stellar
structures. However, trailing secondary bars cannot be explained this
way. Moreover, bars within bars are also observed in SB0 galaxies
whose gas mass fraction is generally very low.
 
\noindent 
5) The best way to reconcile theory and observations consists in
postulating that the secondary stellar bar rotates faster than the
primary stellar bar as suggested by Pfenniger \& Norman
(1990). Friedli \& Martinet (1993) have demonstrated the viability of
such systems which form through a dynamical decoupling between the
central and outer parts of the galaxy. This decoupling is essentially
made possible by the primary bar-driven gas fueling.  Tagger \etal\
(1987) have invoked nonlinear mode coupling to explain multiple
pattern speeds of bar and spiral arms.
 
\noindent 
6) In order to explain the fueling of AGN's, Shlosman \etal\ (1989)
have invoked the formation of a small-scale gaseous bar within a
large-scale primary bar. Simulations by Heller \& Shlosman (1994) have
shown that this process is transient since fragmentation and dynamical
friction quickly dissolve the gaseous bar. So, if this process is to
be very effective in powering AGN's, it is not obvious how to link it
to the existence of persistent misaligned secondary stellar bars.
 
\titleb{Aligned secondary bars} 
Two explanations can be given in order to explain secondary aligned 
stellar bars (i.e. $\theta_d \!=\! 0\degr$) within primary stellar 
bars. This is observed in NGC~1326 after deprojection (Paper I) as 
well as in NGC~4314 (Benedict \etal\ 1993) and in NGC~4321 (Knapen 
\etal\ 1995). 
 
\noindent 
1) There is a single pattern speed associated with a single bar with
two ILR's. The secondary bar is made of orbits trapped by the $x_1$
periodic orbit family inside the inner ILR whereas the primary bar is
supported by the $x_1$ outside the outer ILR to close to the
CR. Between the two ILR's, the $x_2$ dominates, gaseous orbit
crossings are present and spot- or ring-like structures are likely to
be formed there.
 
\noindent 
2) There are two different pattern speeds associated with the two bars 
(see point 5) in Sect.~5.1) and the system is observed in that 
peculiar configuration. 
 
\titleb{Twisted isophotes} 
In order to explain twisted stellar isophotes (i.e. $\theta_d 
\!=\! \theta_d(R)$) within primary stellar bars, the following 
explanations can be put forward. 
 
\noindent 
1) The observed twists come from projection effects: 
 
\noindent 
-- On a triaxial bulge.  This is the most likely explanation for 
early-type galaxies (e.g. NGC~2950).  Note that a triaxial bulge or a 
secondary (thick) bar could in fact be physically indistinguishable. 
 
\noindent 
-- On a thin bar with strongly varying ellipticity with radius. This 
can explain many of the weak \PA\ variations ($\la 10\degr$) observed 
along some primary bars (e.g. NGC~5850). 
 
\noindent 
2) The twists are due to specific central patterns of star formation 
or dust absorption.  In particular, intense and irregular rings of 
star formation can produce isophote twists even in the K-band 
(e.g. NGC~6951).  Similarly, the absorption in the near-IR can remain 
high in some very dusty galaxies (e.g. NGC~1097). However, in general 
this explanation can be dismissed when such twists are observed in the 
near-IR as well. 
 
\noindent 
3) The twists are generated by spiral arms. This generally occurs at 
the end of the primary bar but can also be observed at the end of the 
secondary bar (e.g. NGC~1097). 
 
\noindent 
4) The twists are generated by large amounts of gas (see point 4) in 
Sect.~5.2). This could certainly be an adequate explanation for some 
gas-rich late-type galaxies with leading isophote twists. 
 
\titlea{Conclusions} 
The main results of this paper can summarized as follows: 
 
\noindent 
1) In our sample, near-IR images clearly confirm the existence of 
galaxies with multiple triaxial structures, and in particular 
double-barred galaxies.  In the most dust-rich galaxies some central 
triaxial signatures, like isophote twists, could however still be 
absorption artifacts. 
 
\noindent 
2) The K-band values of the structural parameters are thought to 
trace better the real morphology of the bulk of the stellar mass. 
With some marked exceptions, the structural parameters in the K-band 
are similar to the ones in the I-band. In particular in double-barred 
galaxies, the primary bar lengths measured in the K-band are 
systematically greater (at most 13\%) than the ones in the I-band. 
 
\noindent 
3) The various triaxial structures are robust to standard
deprojection. The deprojected angle $\theta_d$ between the two bars
does not take any preferential value (like 0\degr\ or 90\degr). There
are roughly half leading and half trailing secondary bars.
 
\noindent 
4) Numerical models of two nested bars with two different pattern
speeds fairly well reproduce to first order the observed features of
double-barred galaxies (two distinct PA's, maximum ellipticities, and
surface brightness slopes).  The analysis of various 2D projections of
3D numerical simulations indicates that these bar-within-bar features
are generally strongly reduced as soon as $i \ga 45\degr$. Similarly,
small $\theta$ and/or $e_s^{\rm max}$ are not favourable for bringing
double-barred galaxies to the fore.
 
\noindent 
5) For the K-band, double-barred systems (8 galaxies including 
NGC~470) have (projected values): 
\item{--} 
Primary to secondary bar length ratios ranging from 4.0 to 13.4 with a 
mean value of 7.2. 
\item{--} 
Primary to secondary bar luminosity ratios ranging from 1.8 to 7.5 
with a mean value of 3.5. 
\item{--} 
The mean maximum ellipticity is 0.58 for the primary bar. 
\item{--} 
The mean maximum ellipticity is 0.31 for the secondary bar. 
 
\noindent 
6) Our poor and biased sample does not allow us to infer any firm 
percentage or dependence with Hubble type of the galaxies with bars 
within bars.  The following can nevertheless be said: -- In our 
sample, no double-barred galaxies with types later than T=3 have been 
found. So, galaxies with bars within bars are certainly less abundant 
in later than earlier types. -- In the K-band, the luminosity ratio 
$\gamma$ increases from earlier to later types. 
 
Clearly, misaligned secondary bars or triaxial bulges within primary
bars do exist in nature and open new theoretical prospects and
challenges by accumulating the problems of non-axisymmetric,
time-dependent, non-linear, and dissipative systems.
 
\acknow{This work has been supported by the Swiss National Science  
Foundation (FNRS), the French Academy of Sciences, and in part by the
NSF through Grant AST 91-16442.  We thank the referee, F.~Combes, for
her valuable comments.  DF and HW respectively acknowledge the Steward
Observatory and the Geneva Observatory for their kind hospitality.}
 
\begref{References} 
 
\ref Athanassoula E., 1983, in: Internal Kinematics and Dynamics of 
Galaxies, IAU Symp.~No.~100, ed.~E.~Athanassoula. Reidel, Dordrecht, 
p.~243 
 
\ref Barth A.J., Ho L.C., Filippenko A.V., Sargent W.L.W., 1995, AJ 110, 
1009
 
\ref Baumgart C.W., Peterson C.J., 1986, PASP 98, 56 
 
\ref Benedict G.F., Higdon J.L., Jeffreys W.H., \etal, 1993, AJ 105, 1369 
 
\ref Binney J.J., Gerhard O., 1996, MNRAS (in press) 
 
\ref Boer B., Schulz H., 1993, A\&A 277, 397 
 
\ref Buta R., 1990, ApJ 351, 62

\ref Buta R., Crocker D.A., 1993, AJ 105, 1344 
 
\ref Combes F., 1994, in: Mass-Transfer Induced Activity in Galaxies, 
ed.~I.~Shlosman. Cambridge University Press, Cambridge, p.~170 
 
\ref de Jong R.S., van der Kruit P.C., 1994, A\&AS 106, 451 
 
\ref de Vaucouleurs G., 1974, in: Formation of Galaxies, IAU Symp.~No.~58, 
ed.~J.R.~Shakeshaft. Reidel, Dordrecht, p.~335 
 
\ref Elias J.H., Frogel J.A., Matthews K., Neugebauer G., 1982, AJ 87, 1029 
 
\ref Friedli D., 1996, in: Barred Galaxies, IAU Coll.~No~157, eds.~R.~Buta 
\etal. ASP Conference Series (in press) 
 
\ref Friedli D., Benz W., 1995, A\&A 301, 649
 
\ref Friedli D., Martinet L., 1992, in: Physics of Nearby Galaxies. Nature 
or Nurture?, eds.~T.X.~Thuan \etal.  Editions Fronti\`eres, 
Gif-sur-Yvette, p.~527 
 
\ref Friedli D., Martinet L., 1993, A\&A 277, 27 
 
\ref Garcia G\`omez C., Athanassoula E., 1991, A\&AS 89, 159 
 
\ref Heller C.H., Shlosman I., 1994, ApJ 424, 84 
 
\ref Hummel E., van der Hulst J.M., Keel W.C., 1987, A\&A 172, 32 
 
\ref Jarvis B., Dubath P., Martinet L., Bacon R., 1988, A\&AS 74, 513 
 
\ref Knapen J.H., Beckman J.E., Shlosman I., \etal, 1995, ApJ 443, L73 
 
\ref Kormendy J.J., 1979, ApJ 227, 714 
 
\ref Kormendy J.J., 1982a, ApJ 257, 75 
 
\ref Kormendy J.J., 1982b, in: Morphology and Dynamics of Galaxies, 
12th Advanced Course of the SSAA, eds.~L.~Martinet, M.~Mayor. Geneva 
Observatory, Geneva, p.~113 
 
\ref Moorwood A., Finger G., Biereichel P., \etal, 1992, The Messenger 69, 
61 
 
\ref Ondrechen M.P., van der Hulst J.M., Hummel E., 1989, ApJ 342, 39 
 
\ref Pfenniger D., Norman C., 1990, ApJ 363, 391 
 
\ref Quillen A.C., Frogel J.A., Kuchinski L.E., Terndrup D.M., 1995, AJ 
110, 156 
 
\ref Rhoads J.E., 1996, ApJ (submitted) 
 
\ref Rix H.-W., Rieke M.J., 1993, ApJ 418, 123 
 
\ref Saikia D.J., Pedlar A., Unger S.W., Axon D.J., 1994, MNRAS 270, 46 
 
\ref Sellwood, J., Wilkinson, A., 1993, Rep. Prog. Phys. 56, 173 
 
\ref Shaw M.A., Combes F., Axon D.J., Wright G.S., 1993, A\&A 273, 31 
 
\ref Shaw M.A., Axon D.J., Probst R., Gatley I., 1995, MNRAS 274, 369 
 
\ref Shlosman I., Frank J., Begelman M.C., 1989, Nat 338, 45 
 
\ref Tagger M., Sygnet J.F., Athanassoula E., Pellat R., 1987, ApJ 318, 
L43 
 
\ref Terndrup D.M., Davies R.L., Frogel J.A., DePoy D.L., Wells L.A., 1994, 
ApJ 432, 518 
 
\ref Telesco C.M., Dressler L.L., Wolstencroft R.D., 1993, ApJ 414, 120 
 
\ref van Moorsel G.A., 1982, A\&A 107, 66 

\ref Wakamatsu K., Nishida M.T., 1980, PASJ 32, 389
 
\ref Wozniak H., Friedli D., Martinet L., Martin P., Bratschi P., 1995, 
A\&AS 111, 115 (Paper I) 
 
\endref 
\begfigwid 18cm 
\nomps{} 
\figure{\FPROF}{ 
For each galaxy in our sample: 1) Greyscale and contour maps in J-band
(upper left image), and in K-band (upper right image). Isophotes are
spaced by 0.5 mag. except for NGC~1097 whose isophotes are separated
by 0.25 mag. 2) Plots of surface brightness $\mu$, ellipticity $e$ and
position-angle \PA\ profiles as a function of semi-major axis of the
fitted ellipse for JHK-filters from left to right (lower left
frame). Each point represents a fitted ellipse.  3) Greyscale of the
J--K colour map (lower right image).  Black is the bluest, white the
reddest.  J--H colour map is displayed for NGC~4340 (K-image
saturated) and NGC~6782 (K-image not photometric); H--K colour map is
shown for NGC~5905 (no J-image). For NGC~1097, only the JHK-images are
displayed (photometry not reliable) }
\endfig 

\bye